\documentclass[useAMS,usenatbib]{mnras}

\usepackage[T1]{fontenc}
\usepackage{ae,aecompl}
\usepackage{graphicx}	
\usepackage{amsmath}	
\usepackage{amssymb}	
\usepackage{subfig}
\usepackage{gensymb}
\graphicspath{ {images/} }

\def\aap{Astronomy \& Astrophysics}
\def\apj{The Astrophysical Journal}
\def\mnras{Monthly Notices of the Royal Astronomical Society}
\def\apjs{The Astrophysical Journal Supplement Series}

\newcommand{\vk}{{\mathbf{k}}}
\newcommand{\vy}{{\mathbf{y}}}
\newcommand{\vx}{\mathbf{x}}
\newcommand{\vr}{{\mathbf{r}}}
\newcommand{\dFT}[1]{\frac{\ud^2 #1}{2\pi}\,}

\newcommand{\Cgltwo}{A_{2}}
\newcommand{\ud}{{\text{d}}}

\newcommand{\valpha}{{\boldsymbol{\alpha}}}
\newcommand{\dalpha}{\delta{\boldsymbol{\alpha}}}
\newcommand{\sdalpha}{\delta\alpha}

\newcommand{\grad}{\nabla}

\newcommand{\half}{\frac{1}{2}}
\newcommand{\dkappa}{\delta \kappa}

\title[Dark matter power spectra]{The inner mass power spectrum of galaxies using strong gravitational lensing: beyond linear approximation}

\author[Chatterjee S. et al.]{Saikat Chatterjee,$^{1}$\thanks{E-mail: saikat@astro.rug.nl} L\'{e}on V. E. Koopmans$^{1}$\vspace{0.2cm} \\
$^{1}$Kapteyn Astronomical Institute, University of Groningen, Postbus 800, 9700 AV, Groningen, The Netherlands\\
}

\date{Accepted . Received ; in original form }

\pubyear{2017}

\begin{document}
\label{firstpage}
\pagerange{\pageref{firstpage}--\pageref{lastpage}}
\maketitle

\begin{abstract}
In the last decade the detection of individual massive dark matter sub-halos has been possible using potential correction formalism in strong gravitational lens imaging. Here we propose a statistical formalism to relate strong gravitational lens surface brightness anomalies to the lens potential fluctuations arising from dark matter distribution in the lens galaxy. We consider these fluctuations as a Gaussian random field in addition to the unperturbed smooth lens model. This is very similar to weak lensing formalism and we show that in this way we can measure the power spectrum of these perturbations to the potential. We test the method by applying it to simulated mock lenses of different geometries and by performing an MCMC analysis of the theoretical power spectra. This method can measure density fluctuations in early type galaxies on scales of 1-10 kpc at typical rms-levels of a percent, using a single lens system observed with the {\sl Hubble Space Telescope} with typical signal-to-noise ratios obtained in a single orbit. 
\end{abstract}


\section{Introduction}

According to the Einstein's General Theory of Relativity, light rays (null geodesics) get deflected due to the presence of gravitating objects, a phenomenon called gravitational lensing. In the regime of strong lensing, massive and large cosmic bodies, such as galaxies, can bend light rays coming from a single astrophysical source such that  multiple images of the source are formed \citep[e.g.][]{narayan, kochanek}. By measuring the redshift of the source and the lens galaxy and by analyzing the relative angular positions of the lensed images, their distortions and surface brightness fluctuations, we can put constraints on the mass power spectra of the foreground lens galaxy \citep{cohn, keeton1, schneider1}. Physical processes during the evolution of galaxy, such as accretion, stellar-driven winds, mergers, collapse, feedback from quasars, all have significant roles in shaping a galaxy's mass distribution \citep[e.g.]{somerville}, by studying the mass power-spectrum of galaxies we can gain insight in different galaxy formation scenarios \citep{Rusin, kochanek1}. 
Besides, the smooth matter component, in recent years it has been possible to accurately measure the mass distribution in foreground lens galaxies and also individual dark matter subhalos (down to $\sim 10^8 M_{\odot}$), using strong lensed images (e.g. \citealt{hezaveh1}, \citealt{hezaveh2}, \citealt{fadely}, \citealt{vegetti2}, \citealt{vegetti1}, \citealt{keeton2}).

In this paper we develop a statistical description to relate the mass power-spectrum of lens galaxies, in particular small-scale potential fluctuations in the lens plane, to the statistics of the surface brightness fluctuations in the image plane. We assume that these lens-potential fluctuations can be treated statistically as a Gaussian random field. If the differential deflection of the photon bundles due to these potential perturbations are small, we can reduce their effect to that of weak lensing. We subsequently show that their effect on the power-spectrum can be captured in to a  {\sl Structure Function}, describing the ensemble average of the square of relative deflections between two points in the deflection field as function of their separation. 

An analogous statistical method has been used before in weak gravitational lensing of the Cosmic Microwave Background radiation (e.g., \citealt{lewis}, \citealt{challinor}), although we have further generalized it to the case of strong galaxy-galaxy lensing but without any assumptions of linear first order approximation in the formalism which goes beyond the current statistical methodology available in the literature (see 2012 BSc thesis of Sander Bus\footnote{https://www.astro.rug.nl/opleidingsinstituut/reports/bachelor/}; \citealt{hezaveh}). We present a detailed two-point correlation-function and power-spectrum analysis and we verify the theory with several representative simulated lens systems (i.e. ring, quad and arc).

The paper is organized as follows. In section \ref{theory} we introduce a two point correlation analysis of residual surface brightness fluctuations of the lensed images, after a smooth-model subtraction, and statistically relate this to the power spectrum of residual fluctuations in the lens potential which are not part of the smooth lens model. In section \ref{method} we describe the steps we followed to test the proposed theory on a set of simulated mock lenses. In section \ref{discuss} we end the paper with conclusions and future plans. 

\section{Theory \& Assumptions}\label{theory}

In this section we explain the foundation behind the statistical method to measure the power spectrum of the gravitational-lens potential. The initial assumption is that, to first order, we can model the surface brightness of the observed lensed images using a smooth lens potential (in this paper we illustrate this by using a non-singular Isothermal Ellipsoid, see \citet{kormann}, but any smooth model will do). Subtracting that smooth model from the original images leaves surface-brightness residuals, which we subsequently assume to arise from small potential perturbations in the lens plane. 
We make a number of upfront assumptions in this paper: 
\begin{itemize}
\item[(i)] We assume that there is negligible covariance between the smooth lens mass model parameters and the small-scale potential perturbations. This assumption might break down on the largest scales, but it likely very accurate on scales well below the Einstein radius. 
\item[(ii)] We assume there is negligible covariance between intrinsic source-brightness fluctuations and induced residual image brightness fluctuations as a result of the lens-potential fluctuations. We think this is justified, to some level, because the source brightness distribution is over-constrained thanks to the multiplicity of the lensed images.  
\item[(iii)] We assume that all lens perturbations are Gaussian random fields. For {\sl low mass subhaloes} the Gaussianity assumption holds extremely well as long as the number of the stochastically distributed subhaloes is more than a few within the area in which the power spectra are estimated \footnote{We have verified, by generating a large number of realizations, that the probability density function (PDF) of $N$ number of Poisson distributed particles over an ensemble closely resembles as a Gaussian PDF as long as $N \gtrapprox 4 $. This follows from central limit theorem.}. But if the lens potential is dominated by a {\sl very few massive structures} then this assumptions breaks down and then one has to model them by potential correction formalism (\citealt{vegetti2}). 

\end{itemize}
We will strictly assume these in the current paper but in several follow-up papers, we investigate each of these assumptions in greater detail.

\subsection{Lens potential and surface brightness}

We start with the principle of conservation of surface brightness in lensing. If we denote the surface brightness of source and image by $S(\vy)$ and $I(\vx)$ respectively, then we have
\begin{equation}\label{CSB}
S(\vy) = I(\vx),
\end{equation}
where $\vy$ and $\vx$ are coordinates in the source and in the lens plane, respectively. Here we neglect the PSF and noise in our analysis, which is taken care of in the next section. The lens equation maps points from the image plane to the source plane in the following way
\begin{equation}
\vy(\vx)= \vx - \nabla \psi_0(\vx) - \nabla \delta \psi(\vx),
\end{equation}
where $\psi_0(\vx)$ and $\delta \psi(\vx)$ are the potentials for the smooth lens model and perturbations, respectively. Combining this non-linear lens equation and Eqn.\ref{CSB} gives us 
\begin{equation}
S(\vy)=S(\vx- \valpha - \dalpha) = I(\vx),
\end{equation}
where we have denoted the deflection angle due to smooth model as $\valpha = \nabla \psi_0 $ and deflection due to potential perturbations as $\dalpha = \nabla \delta \psi$. Assuming that $\psi_0$ and $\delta \psi$ are uncorrelated random fields, the two point correlation function of the lensed images $\xi^{II}(r)$ becomes (for a detailed calculation please see Appendix~\ref{APDPS})
\begin{eqnarray}
\xi^{II}(r) &=& \langle I(\vx) I(\vx') \rangle \nonumber \\
&=& \int \frac{\ud^2 \vk}{(2\pi)^2}\, \left[P^{II}_{\rm{s}}(k) \langle e^{i\vk \cdot (\dalpha' - \dalpha)} \rangle_{\dalpha}\right] e^{i \vk \cdot \vr}
\label{flat_lensed_corr}
\end{eqnarray}
where $P^{II}_{\rm{s}}(k)$ is the power spectrum of the images, when lensed only by $\psi_0$. The angle brackets denote the ensemble average over the stochastic field $\dalpha$. Hence we  assume that we can obtain an unbiased, or sufficiently good, estimator of the true source and smooth part of the potential (i.e. the two upfront assumptions). 

In short, the deflections due to the stochastic potential perturbations act as a multiplicative filter on the unperturbed lens-image power-spectrum. The two point correlation function of the lensed image is then the integral transform of the power spectrum of the smooth model using this filter.
If we further assume $\dalpha$ is a Gaussian random field, then $\vk \cdot (\dalpha' - \dalpha)$ is a Gaussian random variate as well and the expectation value in Eq.~\eqref{flat_lensed_corr} reduces to (see Appendix \ref{APDPS}),
\begin{eqnarray}
\left\langle e^{i\vk \cdot (\dalpha' - \dalpha)} \right\rangle 
=e^{- \frac{1}{2} \left\langle [\vk \cdot (\dalpha'-\dalpha)]^2 \right\rangle}.
\label{expavg}
\end{eqnarray}
We call the term $\left\langle [\vk \cdot (\dalpha'-\dalpha)]^2 \right\rangle$ on the right hand side of the above equation the \textsl{deflection-angle structure function} or the \textsl{deflection-angle transfer function}. We note here that unlike similar functions in the literature, this function depends on the distance $r$ and thus is related to a convolution kernel of the smooth image which depends on scale\footnote{A more physical picture is that, to first-order, only fluctuations in $\delta \psi$ on scales $\lesssim r$ significantly contribute to the de-correlation between two points separated by $r$ in the image plane, or $k \gtrsim 2\pi/r$. Hence as $r$ increases, increasingly more of the power-spectrum at lower $k$-values will contribute to deviations from the expected smooth model between two points.}.
Hence the final result for the lensed correlation function in terms of structure function becomes
\begin{equation}
\xi^{II}(r) = \int \frac{\ud^2 \vk}{(2\pi)^2} \left[ P^{II}_{\rm{s}}(k)  e^{-\frac{1}{2}\langle \left[\vk \cdot (\dalpha' - \dalpha)\right]^2\rangle}\right] e^{i \vk \cdot \vr}.
\label{tpcsg}
\end{equation}
The physical interpretation is that the deflection angle structure function, $\left\langle [\vk \cdot (\dalpha'-\dalpha)]^2 \right\rangle$ acts as a blurring function in the kernel over the smooth model due to the small scale structures present in the lens potential. 
\begin{figure*}
  \centering
{\includegraphics[width=35mm]{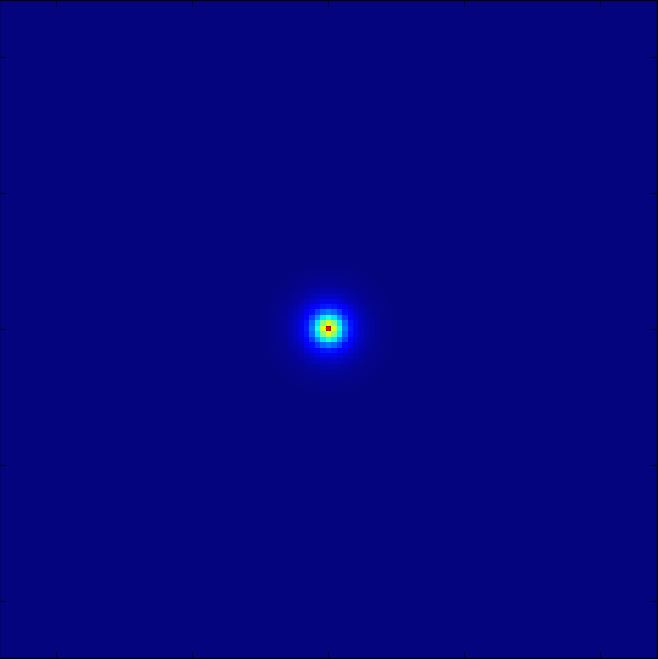}}\qquad
{\includegraphics[width=35mm]{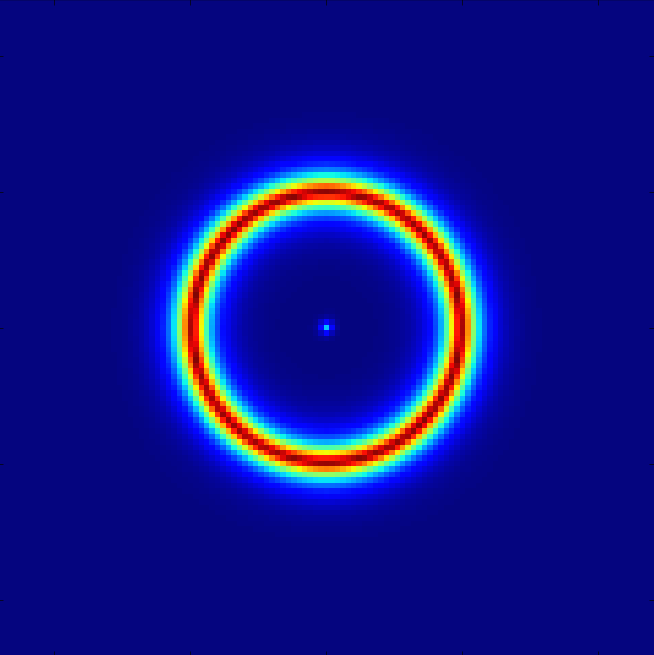}}\qquad
{\includegraphics[width=35mm]{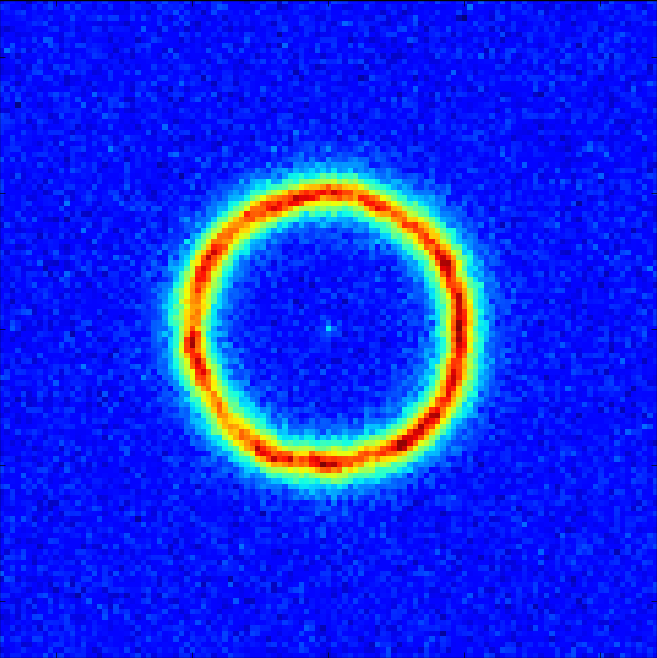}}\qquad
{\includegraphics[width=35mm]{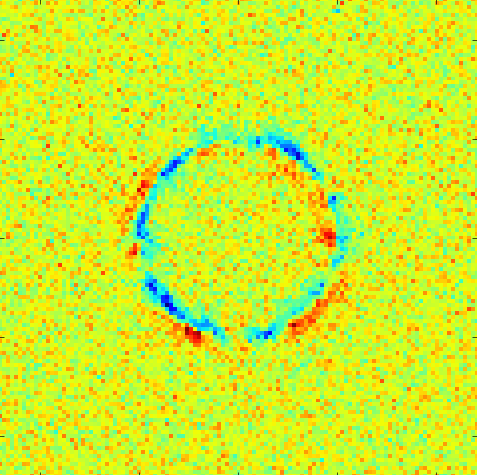}}\\ [15pt]
{\includegraphics[width=35mm]{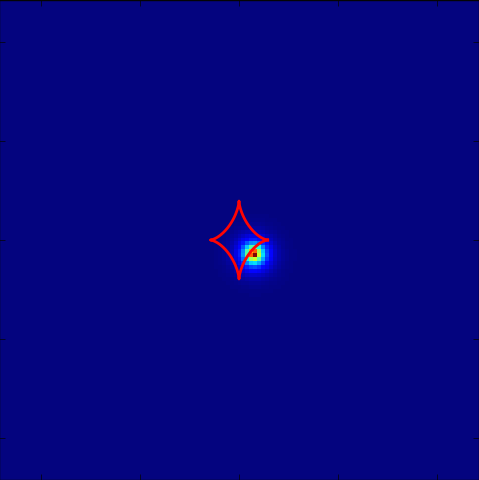}}\qquad
{\includegraphics[width=35mm]{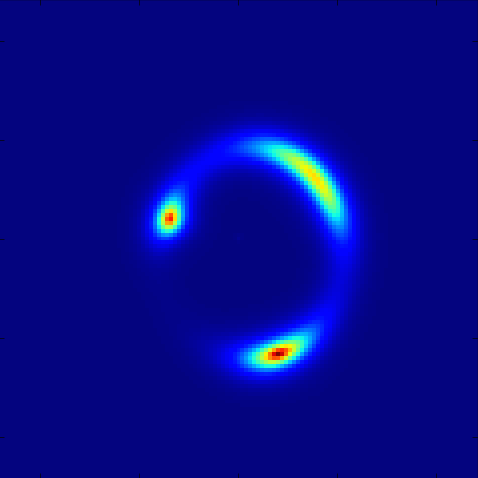}}\qquad
{\includegraphics[width=35mm]{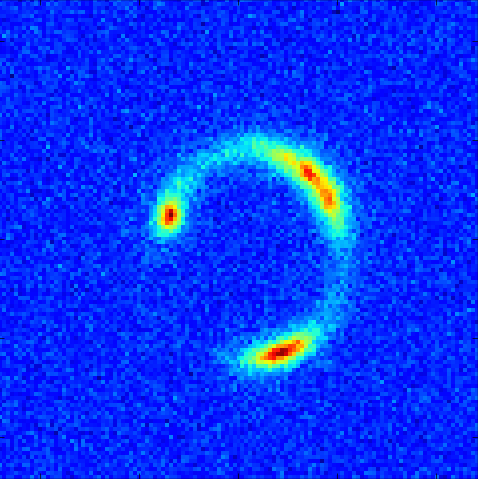}}\qquad
{\includegraphics[width=35mm]{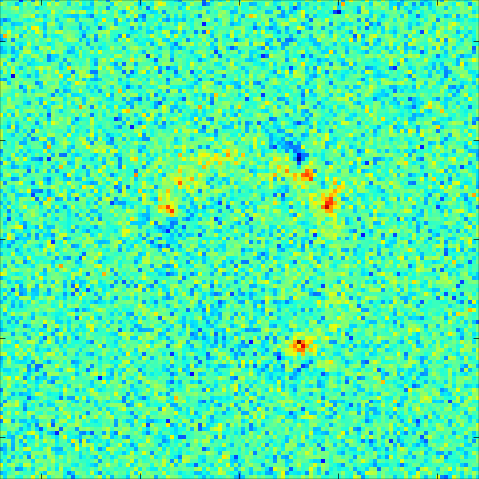}}\\[15pt]
{\includegraphics[width=35mm]{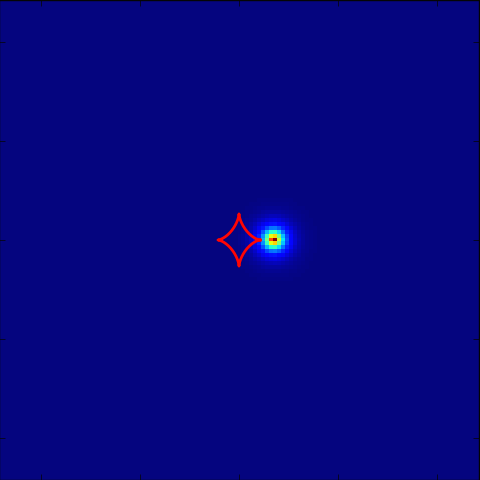}}\qquad
{\includegraphics[width=35mm]{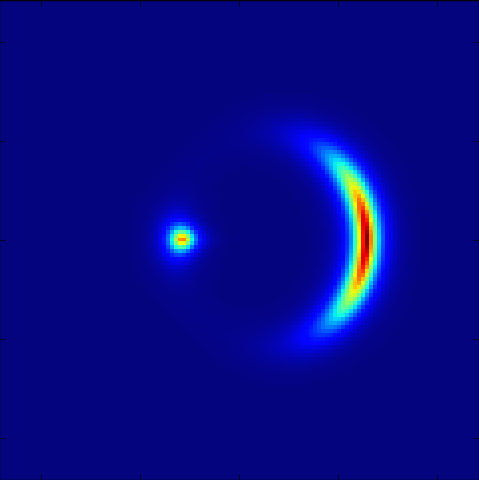}}\qquad
{\includegraphics[width=35mm]{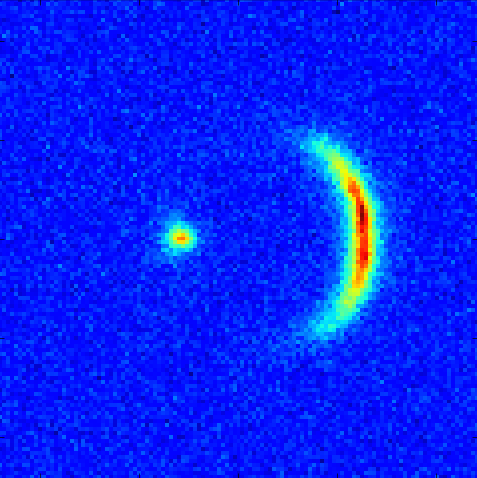}}\qquad
{\includegraphics[width=35mm]{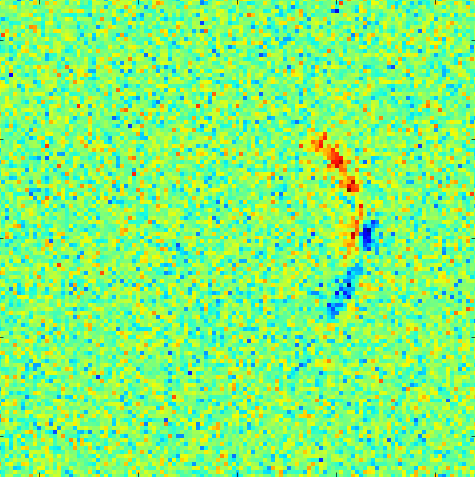}}\\
\caption{Columnwise from left to right: The Sersic source model, the lensed images using a NIE lens model with 1 arsecond Einstein radius (smooth model), the lensed images perturbed by an additive GRF potential with $\sigma^{2}_{\rm{fluct}} = 10^{-3}$ (over the field of view) and a power law of the form $\propto k^{-4}$, respectively. White Gaussian noise is added and the difference between the smooth and perturbed images is shown in the right-most column. The three cases from top to bottom are: Einstein ring,  fold and cusp.}
\label{img}
\end{figure*}

\subsection{The deflection angle structure function}

To compute the deflection angle structure function we compute the correlation matrix:
\begin{equation}
\langle \delta \alpha_i \delta \alpha_j' \rangle = \langle \delta \alpha_i (\mathbf{x}) \delta \alpha_j (\mathbf{x+r})\rangle= \langle \grad_i \delta \psi(\vx) \grad_j \delta \psi(\vx') \rangle
\end{equation}
where $i, j = 1, 2$ denote the components of the deflection $\dalpha$. The correlation matrix of deflections $\langle \delta \alpha_i \delta \alpha_j' \rangle $ can be decomposed into a diagonal component and a off-diagonal component  as in \citet{hezaveh}:
\begin{eqnarray}
\langle \delta \alpha_i \delta \alpha_j' \rangle = A_1(r) \delta_{ij} + A_2(r) \hat{r}_i \hat{r}_j
\end{eqnarray}
where the functions $A_1(r)$ and $A_2(r)$ defined above depend only on the magnitude $r = |\mathbf{r}|$ due to the assumptions of homogeneity and isotropy of the perturbation field. Calculating these functions in terms of the projected surface-mass density ($\delta\kappa$) associated  with the potential perturbations, we find (see Appendix \ref{DASF})
\begin{eqnarray}
A_1(r) &=& \frac{4}{2\pi} \int \frac{\ud k}{k} |\delta\kappa(k)|^2  \frac{J_1(kr)}{kr},  \label{A1} \\
A_2(r) &=& - \frac{4}{2\pi} \int \frac{\ud k}{k} |\delta\kappa(k)|^2  J_2(kr). \label{A2}
\end{eqnarray}
where $A_1(r)$ and $A_2(r)$ represent the correlation in the deflection field between two points separated by a distance $r$, integrated over all $k$ modes, assuming that the power-spectra of the potential or in this case the convergence (i.e.\ $|\delta\kappa(k)|^2$) perturbations are isotropic. We now express the required expectation value in Eq.~\eqref{expavg} in terms of $A_1$ and $A_2$ (a detailed derivation is given in Appendix \ref{DASF}):
\begin{eqnarray}
\left\langle [\vk \cdot (\dalpha'-\dalpha)]^2 \right\rangle
&=&k^2 \Big[ \sigma^2(r) + \cos 2\phi\, \zeta(r) \Big] \nonumber \\
&=& k^2 \sigma^2(r) + ( k_{r_{\parallel}}^2 - k_{r_{\perp}}^2 )\zeta(r) \nonumber \\
\label{flat_expect}
\end{eqnarray}
where $\phi$ is the angle between $\vk$ and $\vr$. We also have resolved $k$ into two components, parallel and perpendicular to $r$. The $r$ appearing in the functions $\sigma^2(r), \zeta(r)$ is the distance between two points in the random field between which we are measuring the correlation. Here we defined the isotropic term as $\sigma^2(r)$,  which is half the variance of the relative deflection of the two points, as follows:
\begin{align}
\sigma^2(r) &=\half\langle (\dalpha-\dalpha')^2\rangle \nonumber \\
&= \half \langle \dalpha \cdot \dalpha + \dalpha' \cdot \dalpha' -2\dalpha \cdot \dalpha' \rangle \nonumber \\
&=  \Big[ \Big( 2 \, A_1(0) + A_2(0) \Big) -  \Big( 2 \, A_1(r) + A_2(r) \Big) \Big] \nonumber \\
&= 2 \Big( A_1(0) - A_1(r) \Big) + \Big( A_2(0) - A_2(r) \Big)
\end{align}
and the function $\zeta(r)$, which determines the anisotropy in the correlation matrix, is defined as follows:
\begin{equation}
\zeta(r) = A_2(0) - A_2(r) .
\end{equation}
If the field is isotropic then $\zeta(r) = 0$ and the structure function in Eqn.~\ref{tpcsg} reduces to a (scale-dependent) Gaussian convolution kernel (Section 4.2.3 in \citealt{lewis}). 

So far all the results from our analysis are without any approximations. But we can also Taylor expand the exponential up to first order to get the following perturbation series,
\begin{eqnarray}\label{tpc_series}
\xi^{II}(r) & \approx &  \xi^{II}_{\rm{s}}(r) - \int  \frac{ k \, \ud k }{2\pi} P^{II}_{\rm{s}}(k) \frac{k^2}{2}\sigma^2(r)J_0(kr) \nonumber \\
& + &  \int \frac{ k \, \ud k }{2\pi} P^{II}_{\rm{s}}(k) \frac{k^2}{2} \zeta(r) J_2(kr),
\end{eqnarray}
where $\xi^{II}_{\rm{res}}(r) =  \xi^{II}(r) - \xi^{II}_{\rm{s}}(r)$ is the correlation function of the residuals which is only a function of the correlation length $r$. The power spectra of the residuals can be obtained from $\xi^{II}(r)_{\mathbf{res}}$ by a Hankel Transform. 

Although any integrable power-spectrum model  can be used, for the sake of simplicity in this paper assume that the power spectrum of lens potential fluctuations follow a power law and hence (from Eq. \ref{dkappa}), convergence fluctuation power spectra in the integral expressions of $A_1(r)$ and $A_2(r)$ in Eq. \ref{A1} and \ref{A2} can also be expressed as a power law. In this case those integrals have analytically exact results in terms of Hypergeometric functions (Appendix \ref{hgf}) which are inserted into the expression of the transfer function and are used in our MCMC fit of the power spectrum (see Section \ref{mcmc}). 
In our subsequent analysis we drop the anisotropy term and only keep the isotropic term in the structure function. 
To compare with the simulations and the real data and to set constraints to the observed power spectrum, we need to incorporate a point spread function (PSF) and noise to our two point correlation formalism. We take these into account in the next section.
\section{Simulations and Results}\label{method}

In this section we summarize the methodology that we followed to test the theory on the simulated mock lenses and discuss the results. First we show how we take into account PSF and noise in our theoretical expression and then we briefly explain how we simulate the mock lenses to test our theory. We test the formalism on three different topologies, a ring, a fold and a cusp, and for each of these cases we consider three different slopes and three different normalizations (i.e.\ set by the variance of the GRF inside the field of the view being simulated) for the power spectrum of the potential fluctuations. 
\subsection{Point Spread Function and Noise}
To take account of the observational effects (e.g., seeing and noise), we first incorporate the effect of a PSF in the two-point correlation analysis. The surface brightness of the unperturbed model becomes,
\begin{equation}
\tilde{I}_{\rm{s}}(\vx) = I_{\rm{s}}(\vx) \otimes \rm{PSF}(\vx).
\end{equation}
So, the observed two-point correlation function of the smooth model will be, 
\begin{equation}
\langle \tilde{I}_{\rm{s}}(\vx) \tilde{I}_{\rm{s}}(\vx') \rangle =  \xi^{II}_{\rm{s}}(|\vx - \vx'|) \otimes \xi^{\rm{PSF}}(|\vx - \vx'|) 
\end{equation}
And, the power spectrum becomes,
\begin{align}
\langle \tilde{I}_{\rm{s}}(\vk) \tilde{I}_{\rm{s}}^*(\vk') \rangle = P^{II}_{\rm{s}} (k) \cdot P^{\rm{PSF}}(k) 
\end{align}
where we have used the convolution theorem.
In our simulation we have added Gaussian noise with a flat power spectrum to the lensed images. We assume that the noise realizations and the images are independent random fields, the power spectrum of sum becomes the sum of their power spectra, 
\begin{align}
\tilde{P}_{\rm{s}}^{II} (k)= P_{\rm{s}}^{II} (k) \cdot P^{\rm{PSF}}(k) + P^{\rm{n}}(k) 
\end{align}
where $P^{\rm{n}}(k)$ is the power spectrum of the noise. Thus, the theoretical expression of two point correlation function becomes,
\begin{eqnarray}\label{thpsf}
\xi^{II}_{\mathbf{\, obs}}(r) &=& \int \frac{\ud^2 \vk}{(2\pi)^2}\, \tilde{P}_{\rm{s}}^{II}(k)  e^{i \vk \cdot \vr}
\langle e^{i\vk \cdot (\dalpha' - \dalpha)} \rangle_{\dalpha}  \\
&=& \int \frac{\ud^2 \vk}{(2\pi)^2}\, \left[ {P_{\rm{s}}^{II}} (k) \cdot T(k) + P^{\rm{n}}(k) \right] e^{i \vk \cdot \vr},\nonumber
\end{eqnarray}
with the modified transfer function
\begin{equation}
T(k) \equiv P ^{\rm{PSF}}(k) \cdot e^{-\frac{1}{2}\langle \left[\vk \cdot (\dalpha' - \dalpha)\right]^2\rangle}.
\end{equation}

\subsection{Lens model and Potential perturbations}

To test the above theory we simulate the lensed images of Sersic (\citet{sersic1}, \citet{sersic2}) sources by a Non-singular Isothermal Ellipsoid (NIE) (\cite{kormann}) lens which we call as smooth model. The values of the lens and source parameters are given in Table \ref{parameters}. All the simulated images have roughly the same resolution as that of HST in the F390W-band.

\begin{table}
\caption{The lens and source parameters chosen for simulating mock lenses of 121x121 pixels in $4.84''$x $4.84''$ field of view.}
	\begin{center}
		\begin{tabular}{l c c}
Parameter              & Value & Unit \\
\hline
\multicolumn{3}{c}{Lens (NIE) (ring, fold, cusp cases)}\\
\hline
x-coordinate         & $0.0$ & arcsecond\\
y-coordinate         & $0.0$ & arcsecond\\
Einstein radius      & $1.0$ & arcsecond\\
Axis ratio           & 0.99, 0.6, 0.7 & -\\
Major-axis angle     & $0.0$ & degree\\
External shear       & 0.0  & -\\
External-shear angle & $0.0$ & degree\\
\hline
\multicolumn{3}{c}{Source (S\'ersic) (ring, fold, cusp cases)}\\
\hline
x-coordinate         & $0.0, 0.15, 0.35$ & arcsecond\\
y-coordinate         & $0.0, 0.15, 0.0$ & arcsecond\\
Effective radius      & $0.1, 0.07, 0.08$  & arcsecond\\
Axis ratio           & 0.99 & -\\
Major-axis angle     & $45$ & degree\\
S\'ersic index       & 2 & -\\
\hline
		\end{tabular}
		\label{parameters}
	\end{center}
\end{table}

We perturb the lens potential by a simulated Gaussian random field (GRF) potential with a power-law power-spectrum of the form 
\begin{equation}\label{grfps}
P^{\delta \psi} (k) = {\cal A} \cdot k^{-\beta}
\end{equation}
where the amplitude ${\cal A}$ in the power law is determined using Parseval's theorem, which is related to the variance of the GRF potential fluctuations inside the image $\sigma^{2}_{\rm{fluct}}$ via the normalization factor
\begin{equation}
{\cal A} = \frac{\sigma^{2}_{\rm{fluct}} N_{\rm{pix}}^2}{2 \sum k^{-\beta}},
\end{equation}
where the sum is over all $k-$values where $k = \sqrt {k_x^2 + k_y^2} $ calculated on the Fourier grid of size 121$\times$121 in our simulations. In this case a DFT of one random realization of the above power spectrum leads to a GRF potential field with a variance of $\sigma^{2}_{\rm{fluct}}$. Now because of the symmetry, we only create half of the grid and the rest is generated from the complex conjugate of it. However in Fourier space, a point and it's complex conjugate aren't independent of each other which results in an increase in variance by a factor of two which is taken care of in the denominator of the normalization factor above. 
We choose the power law exponent in the range of $\beta = 4 \sim 6$ and the power spectrum is set to zero at $k=0$ to avoid a non-zero mean value.

\begin{figure*}
  \centering
{\includegraphics[width=92mm]{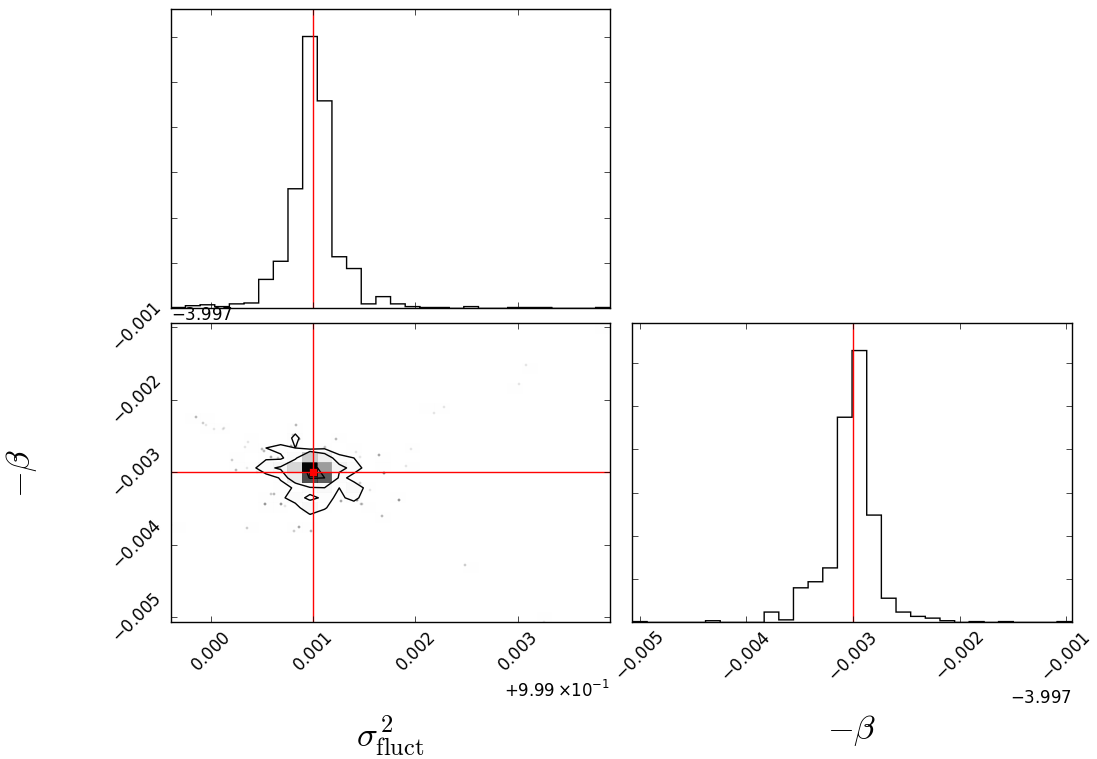}}\quad
{\includegraphics[width=82mm]{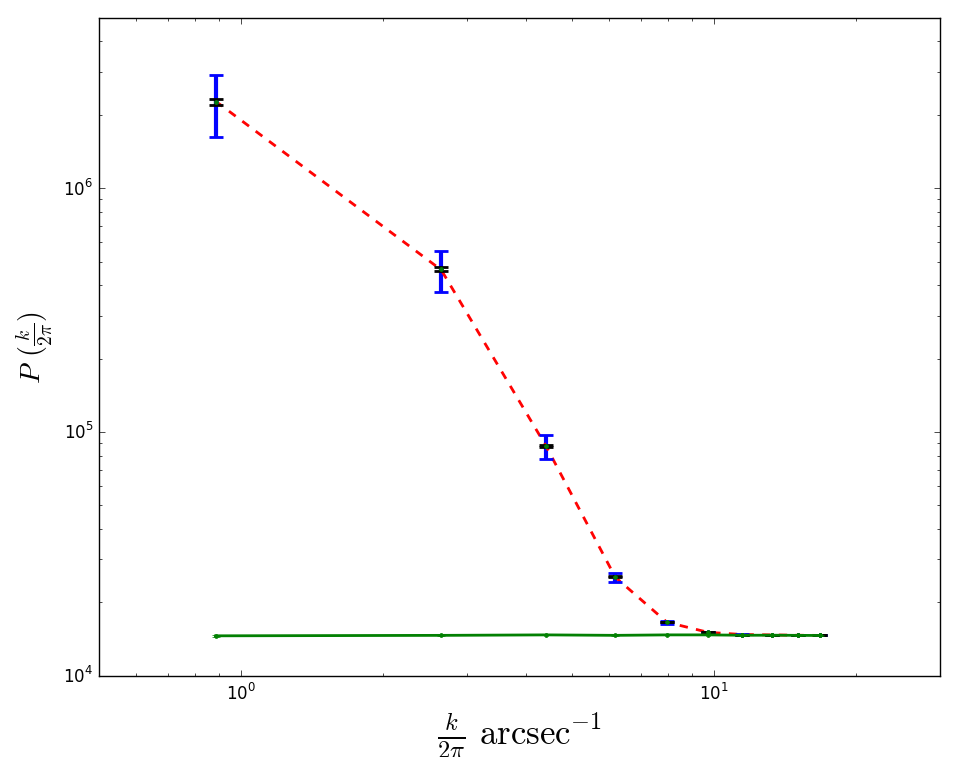}}\\
[8pt]
{\includegraphics[width=92mm]{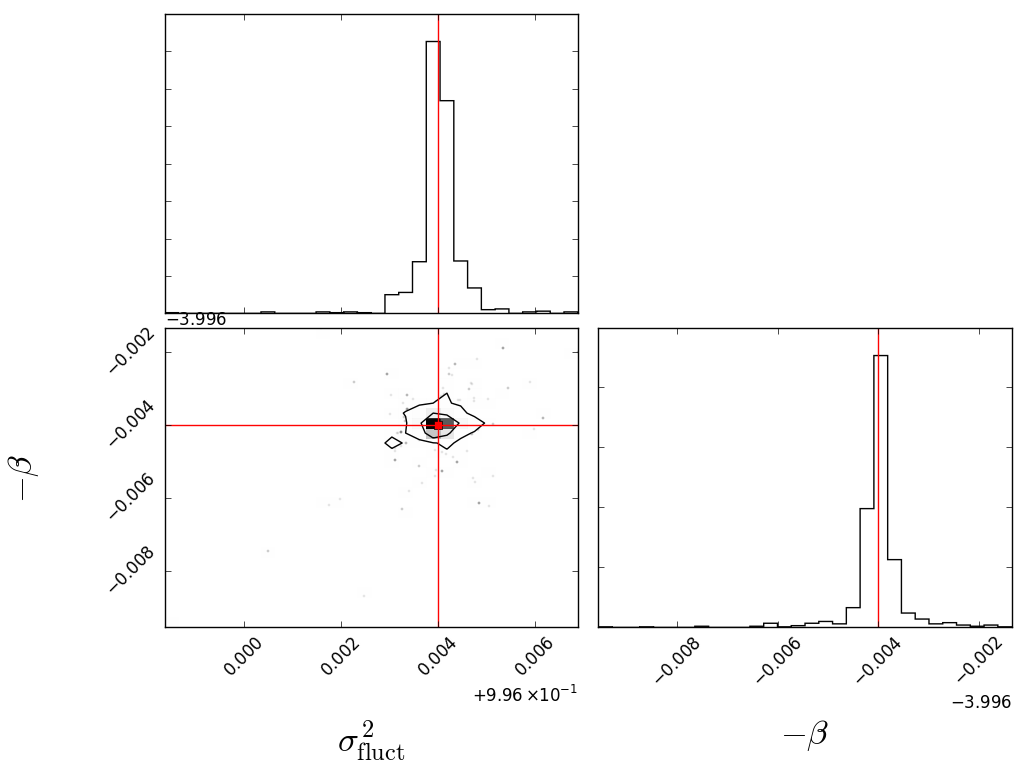}}\quad
{\includegraphics[width=82mm]{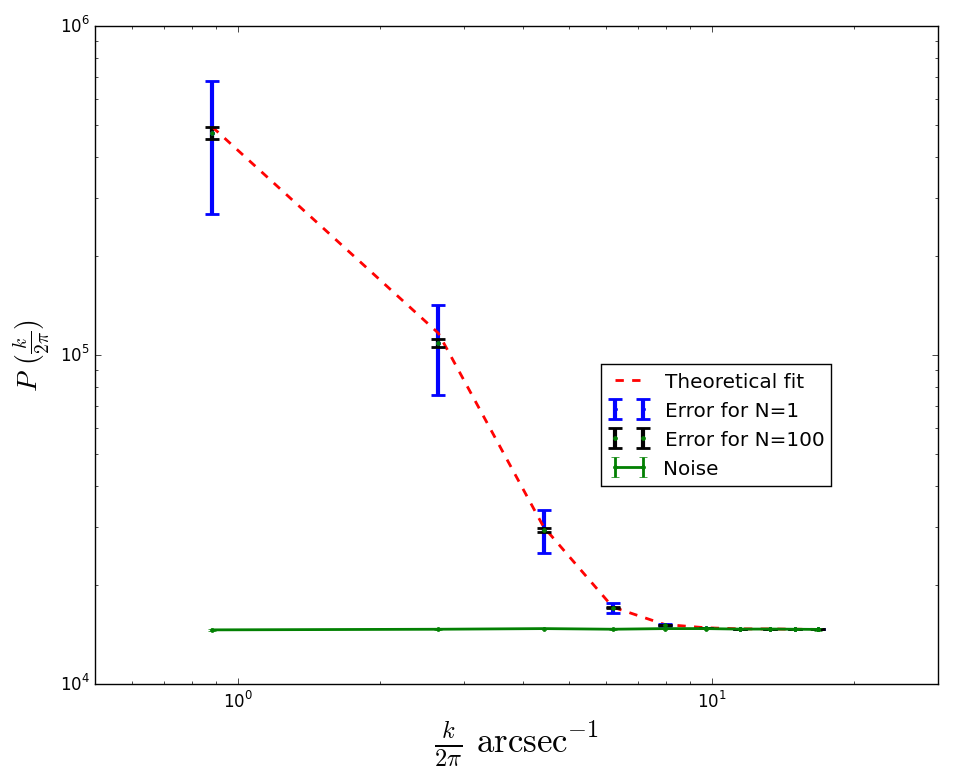}}\\
[8pt]
{\includegraphics[width=92mm]{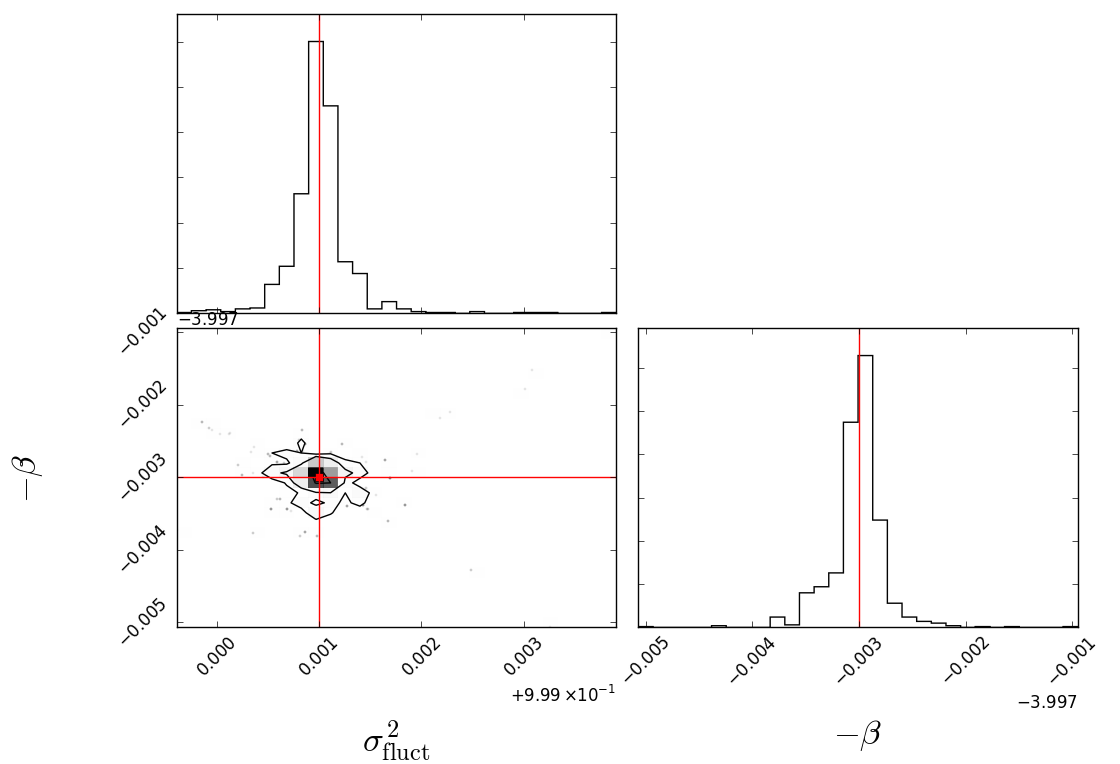}}\quad
{\includegraphics[width=82mm]{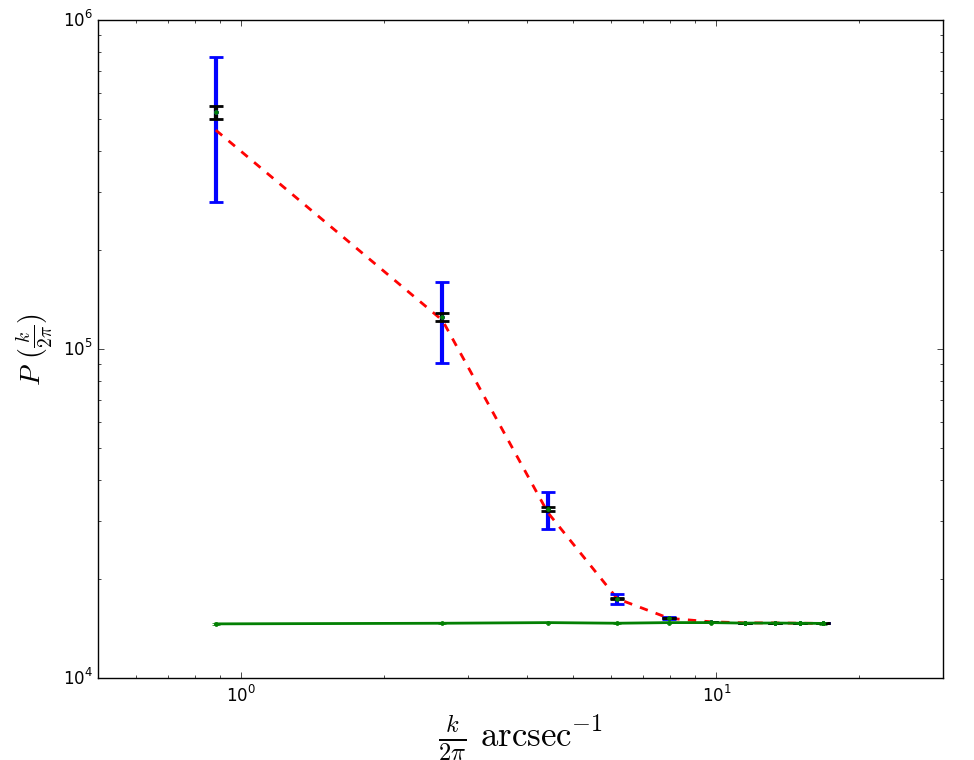}}
\caption{Corner plots (left) for a two-parameter MCMC fit for $\sigma^{2}_{\rm{fluct}}$ and $\beta$ of the GRF potential fluctuations. From top to bottom we show the three different geometries: the ring, fold and cusp. The red lines indicate the input parameters. In the right panels, the red dashed curves are the theoretical power spectra of the surface brightness fluctuations (from which the results are inferred), over-plotted by the recovered mean residual power spectra of 100 realizations (shown in blue with the error bars for a single realization [large]and one hundred realizations [small]) and the white Gaussian noise (green horizontal line) with a variance of 1.0. }
\label{ps}
\end{figure*}
\begin{figure*}
  \centering
{\includegraphics[width=193mm]{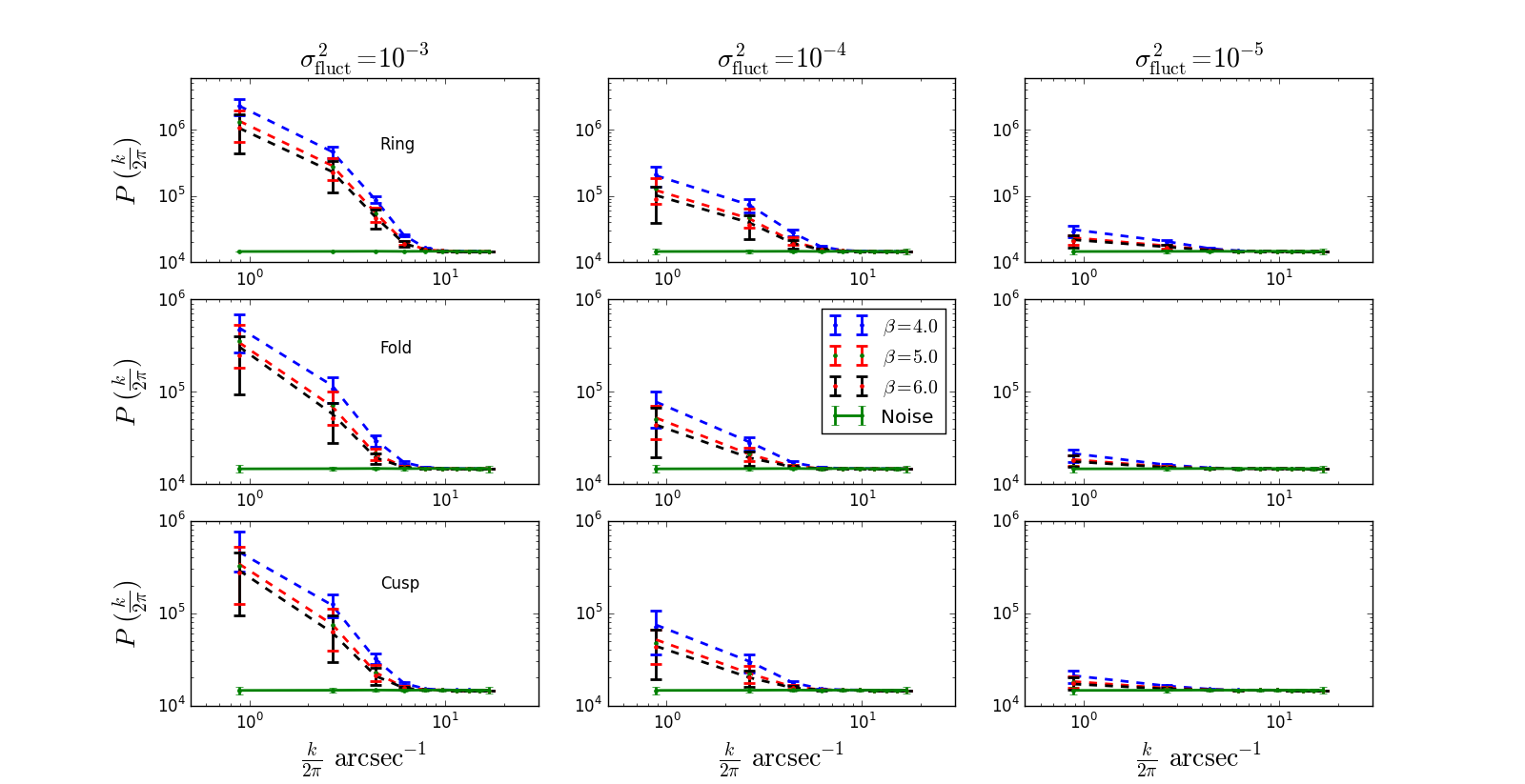}}
\caption{Power spectra of surface-brightness residuals for a set of different variances and slopes of the Gaussian Random field of potential fluctuations. Shown are the three different geometries: ring, fold \& cusp. We notice that the results are not strongly dependent on the geometry, but for a fixed overall variance, the less-steep power-spectra for the GRF yield strong surface-brightness fluctuations.}
\label{grf}
\end{figure*}

We convolve the lensed images with the HST U-band PSF (WFC3 UVIS channel, F390W) obtained using {\tt TinyTim} \citep[][]{Krist:2011kt}. We add white  noise of variance comparable to a typical single-orbit HST image, and we determine the residuals by subtracting the unperturbed smooth model from the perturbed one. Again, we assume that there is no strong covariance between $\delta\psi$ and the parameters of the lens and source models. This likely holds for intermediate $k$-values (see Bayer et al. 2017, in prep.), but not for those comparable to the scale of the lens or very small $k$-values, where they can either be affected by the smooth lens model or by the grid-based source model. In ongoing simulations, and expansion of the theory, this co-variance will be further investigated in the near future. A panel of simulated lensed images for the case of a ring, a fold and a cusp are shown in Figure~\ref{img}, for both the smooth and a perturbed case, as well as their difference. 

\subsection{Realizations}\label{mcmc}

We finally simulate one hundred realizations for each of the three geometries and fit the ensemble-average power spectrum of residuals using Eqn.~\ref{thpsf}. We use a Markov Chain Monte Carlo (MCMC) method to infer the variance and the exponent of the power spectrum of the lens-potential fluctuations that we defined in Eqn. \ref{grfps}. We assume that the power spectrum of smooth lensed image ${P_{\rm{s}}^{II}} (k)$ can be estimated within sufficient accuracy with limited covariance between the source model and potential fluctuations. This assumption has recently been validated by numerical simulations over most angular scales in the lensed images (Bayer, Chatterjee et al., to be submitted). 

\begin{figure*}
  \centering
{\includegraphics[width=145mm]{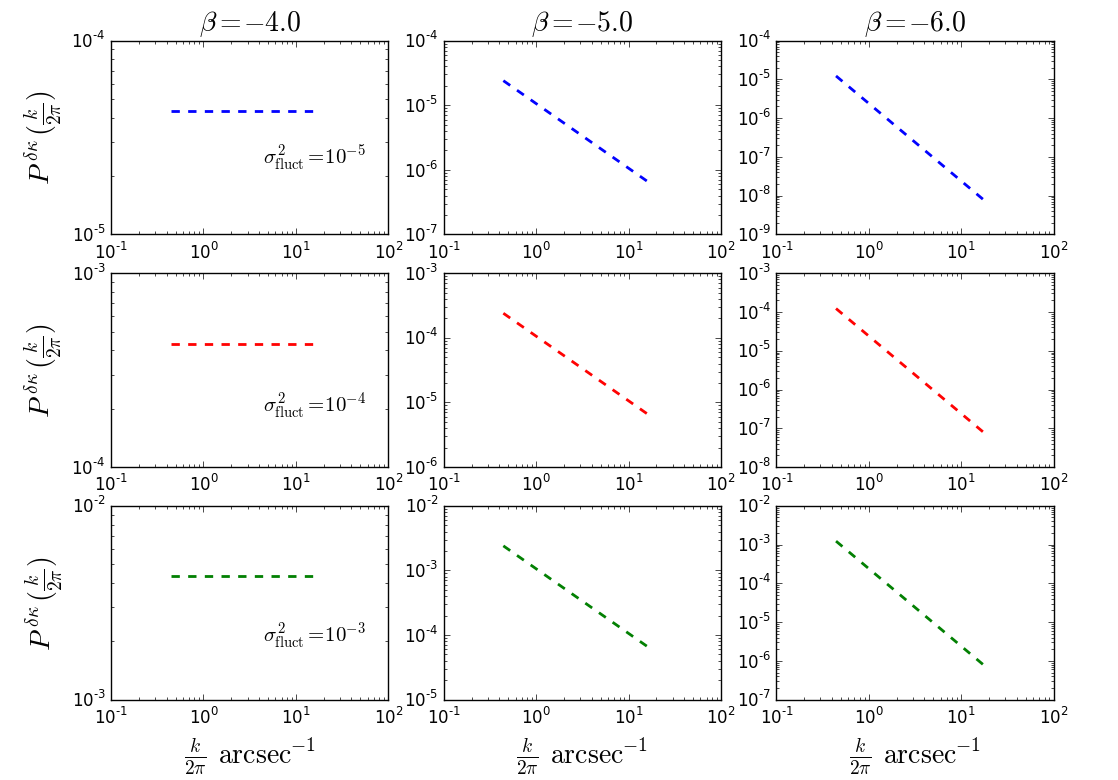}}
\caption{Power spectra of residual convergence map $\delta \kappa$, corresponding to different combinations of variances $\sigma^{2}_{\rm{fluct}}$ and slopes $\beta$ of the lens potential fluctuations, $\delta \psi$. From upper to lower rows (blue, red \& green) $\sigma^{2}_{\rm{fluct}}$ corresponds to $10^{-5}, 10^{-4}$ and $10^{-3}$ and from left to right $\beta$ corresponds to -4.0, -5.0 and -6.0 respectively.}
\label{kappa}
\end{figure*}

\begin{figure*}
  \centering
{\includegraphics[width=145mm]{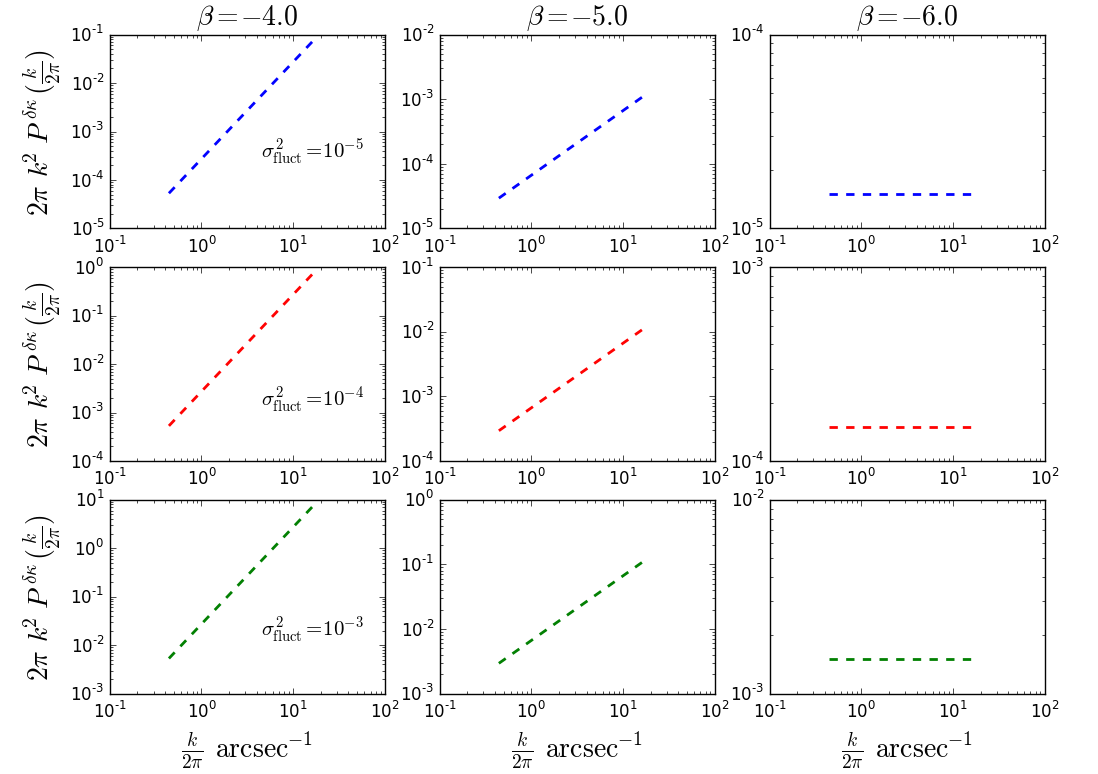}}
\caption{Plots of $2 \pi k^2 P^{\delta \kappa}$ corresponding to the same combinations of variances and slopes of potential fluctuations, $\delta \psi$ as shown in Figure \ref{kappa}.}
\label{kappa_2}
\end{figure*}

In principle we could also fit for the RMS of noise power-spectrum in the likelihood function, but it is not needed at this point (it can often be determined from other parts of the image without lensed images) and we co-add the power spectra of simulated noise and of the residuals. The error (variance on the variance) of the power spectrum is calculated for each bin $j$ via the root mean square deviation from the mean within the ensemble of realizations,
\begin{equation}
{\rm rms}({P_j}) = \left(\sum_{i=1}^{N} (P_{ij} - \langle P \rangle_{j})^2\right)/({N-1})
\end{equation}
where in our case we took $N=100$. This is the error for a single measurement and error for $N$ observations is determined by dividing the error for a single measurement by a factor of $\sqrt{N}$. In reality the $N=100$ lenses will have different sets of lens-model parameters, which we ignore in this paper. We note that the ESA space-mission {\sl Euclid} might discover sufficient numbers of lenses that samples of order one hundred similar geometries could be discovered, although our method also works for ensembles of very different lens geometries. The MCMC corner plots for the three geometries are shown in Figure \ref{ps}. We also show power spectra of surface brightness residuals for different combinations of $\sigma^{2}_{\rm{fluct}}$ and $\beta$  in Figure \ref{grf}. Power spectra of residuals in convergence maps corresponding to those combinations of parameters and their variances are shown in Figure \ref{kappa} and Figure \ref{kappa_2}. Also in Figure \ref{kappa_3} we have shown a comparison of power spectra obtained from numerically calculated convergence by directly applying a Laplacian on the potential map:
\begin{equation}
\delta \kappa = \nabla^2 \delta \psi /2 
\end{equation}
and the theoretical one:
\begin{equation}\label{dkappa}
P^{\delta \kappa} (k) = (2 \pi k)^4 \cdot P^{\delta \psi} (k) /4
\end{equation}
assuming $P^{\delta \psi} (k)$ follows a power law as defined in Eq. \ref{grfps}. The factor $(2 \pi)^4$ in Eq. \ref{dkappa} comes from our definition of Fourier kernel where $k\equiv 1/L$, which differs from the standard cosmological definition via a simple coordinate transformation.

\section{Discussion and Conclusions}\label{discuss}

We have shown that small fluctuations in the gravitational-lens potential, if well-approximated by a (Gaussian) random field, can be treated as a stochastic contribution to the smooth lens model. Assuming further that there is no strong covariance between the smooth lens potential and these lens-potential fluctuations {\sl and} that the surface brightness fluctuations are not affected too much by the inference of the source model, we have developed a  statistical method which can be used to measure the power spectrum of these lens-potential perturbations directly from the power spectrum of the surface brightness fluctuations after subtracting the best smooth lens model. In a forthcoming paper (Bayer et al. 2017, in prep.) we will apply this method to HST images of one particular lens system, more precisely defining (via simulations) to what level these assumption hold.

Quantitatively we have shown that perturbations to the potential or convergence at the percent-level (rms) can be inferred from a single lens system with HST-like images and a typical signal-to-noise ratio in a single orbit. The inference does not strongly depend on the 
geometry of the lens (e.g. ring, fold or cusp), although the ring-geometry seems to show somewhat smaller errors (see Fig.\ref{grf})

In a forthcoming paper (Bayer et al., 2017, in preparation) we will apply this approach to HST data to set limits on the power spectrum of the potential fluctuations around a massive early-type galaxy. Our new method can infer density fluctuations in galaxies on scales of typically 1-10 kpc, in the regime where very little is known about the galaxy (or CDM) power-spectrum. The final goal is to compare these power-spectra to different galaxy-formation scenarios, by applying the method to mock lenses simulated via N-body hydrodynamic simulations. This method will also pave the way for the future (statistical) modeling of hundreds of thousands strong lenses expected to be found from ESA's {\sl Euclid} mission. 
\section*{Acknowledgment}
This work was financially supported by a grant (project number 614.001.206) from the Netherlands Organization for Scientific Research (NWO). S.C. would like to thank Dr. John McKean for helpful and constructive discussions related to observational aspects. 
\bibliographystyle{mnras}
\bibliography{references}

\begin{thebibliography}{}
\makeatletter
\relax
\def\mn@urlcharsother{\let\do\@makeother \do\$\do\&\do\#\do\^\do\_\do\%\do\~}
\def\mn@doi{\begingroup\mn@urlcharsother \@ifnextchar [ {\mn@doi@}
  {\mn@doi@[]}}
\def\mn@doi@[#1]#2{\def\@tempa{#1}\ifx\@tempa\@empty \href
  {http://dx.doi.org/#2} {doi:#2}\else \href {http://dx.doi.org/#2} {#1}\fi
  \endgroup}
\def\mn@eprint#1#2{\mn@eprint@#1:#2::\@nil}
\def\mn@eprint@arXiv#1{\href {http://arxiv.org/abs/#1} {{\tt arXiv:#1}}}
\def\mn@eprint@dblp#1{\href {http://dblp.uni-trier.de/rec/bibtex/#1.xml}
  {dblp:#1}}
\def\mn@eprint@#1:#2:#3:#4\@nil{\def\@tempa {#1}\def\@tempb {#2}\def\@tempc
  {#3}\ifx \@tempc \@empty \let \@tempc \@tempb \let \@tempb \@tempa \fi \ifx
  \@tempb \@empty \def\@tempb {arXiv}\fi \@ifundefined
  {mn@eprint@\@tempb}{\@tempb:\@tempc}{\expandafter \expandafter \csname
  mn@eprint@\@tempb\endcsname \expandafter{\@tempc}}}

\bibitem[\protect\citeauthoryear{{Challinor} \& {Lewis}}{{Challinor} \&
  {Lewis}}{2005}]{challinor}
{Challinor} A.,  {Lewis} A.,  2005, \mn@doi [\prd]
  {10.1103/PhysRevD.71.103010}, \href
  {http://adsabs.harvard.edu/abs/2005PhRvD..71j3010C} {71, 103010}

\bibitem[\protect\citeauthoryear{{Cohn}, {Kochanek}, {McLeod}  \&
  {Keeton}}{{Cohn} et~al.}{2001}]{cohn}
{Cohn} J.~D.,  {Kochanek} C.~S.,  {McLeod} B.~A.,   {Keeton} C.~R.,  2001,
  \mn@doi [\apj] {10.1086/321412}, \href
  {http://adsabs.harvard.edu/abs/2001ApJ...554.1216C} {554, 1216}

\bibitem[\protect\citeauthoryear{{Fadely} \& {Keeton}}{{Fadely} \&
  {Keeton}}{2012}]{fadely}
{Fadely} R.,  {Keeton} C.~R.,  2012, \mn@doi [\mnras]
  {10.1111/j.1365-2966.2011.19729.x}, \href
  {http://adsabs.harvard.edu/abs/2012MNRAS.419..936F} {419, 936}

\bibitem[\protect\citeauthoryear{{Hezaveh}, {Dalal}, {Holder}, {Kuhlen},
  {Marrone}, {Murray}  \& {Vieira}}{{Hezaveh} et~al.}{2013}]{hezaveh2}
{Hezaveh} Y.,  {Dalal} N.,  {Holder} G.,  {Kuhlen} M.,  {Marrone} D.,  {Murray}
  N.,   {Vieira} J.,  2013, \mn@doi [\apj] {10.1088/0004-637X/767/1/9}, \href
  {http://adsabs.harvard.edu/abs/2013ApJ...767....9H} {767, 9}

\bibitem[\protect\citeauthoryear{{Hezaveh}, {Dalal}, {Holder}, {Kisner},
  {Kuhlen}  \& {Perreault Levasseur}}{{Hezaveh} et~al.}{2014}]{hezaveh}
{Hezaveh} Y.,  {Dalal} N.,  {Holder} G.,  {Kisner} T.,  {Kuhlen} M.,
  {Perreault Levasseur} L.,  2014, preprint, \href
  {http://adsabs.harvard.edu/abs/2014arXiv1403.2720H} {} (\mn@eprint {arXiv}
  {1403.2720})

\bibitem[\protect\citeauthoryear{{Hezaveh} et~al.,}{{Hezaveh}
  et~al.}{2016}]{hezaveh1}
{Hezaveh} Y.~D.,  et~al., 2016, \mn@doi [\apj] {10.3847/0004-637X/823/1/37},
  \href {http://adsabs.harvard.edu/abs/2016ApJ...823...37H} {823, 37}

\bibitem[\protect\citeauthoryear{{Keeton}}{{Keeton}}{2001}]{keeton1}
{Keeton} C.~R.,  2001, ArXiv Astrophysics e-prints, \href
  {http://adsabs.harvard.edu/abs/2001astro.ph..2341K} {}

\bibitem[\protect\citeauthoryear{{Keeton} \& {Madau}}{{Keeton} \&
  {Madau}}{2001}]{keeton2}
{Keeton} C.~R.,  {Madau} P.,  2001, \mn@doi [\apjl] {10.1086/319136}, \href
  {http://adsabs.harvard.edu/abs/2001ApJ...549L..25K} {549, L25}

\bibitem[\protect\citeauthoryear{{Kochanek} et~al.,}{{Kochanek}
  et~al.}{2000}]{kochanek1}
{Kochanek} C.~S.,  et~al., 2000, \mn@doi [\apj] {10.1086/317074}, \href
  {http://adsabs.harvard.edu/abs/2000ApJ...543..131K} {543, 131}

\bibitem[\protect\citeauthoryear{{Kormann}, {Schneider}  \&
  {Bartelmann}}{{Kormann} et~al.}{1994}]{kormann}
{Kormann} R.,  {Schneider} P.,   {Bartelmann} M.,  1994, \aap, \href
  {http://adsabs.harvard.edu/abs/1994A%26A...284..285K} {284, 285}

\bibitem[\protect\citeauthoryear{Krist, Hook  \& Stoehr}{Krist
  et~al.}{2011}]{Krist:2011kt}
Krist J.~E.,  Hook R.~N.,   Stoehr F.,  2011, SPIE Optical {\ldots}, 8127,
  81270J

\bibitem[\protect\citeauthoryear{{Lewis} \& {Challinor}}{{Lewis} \&
  {Challinor}}{2006}]{lewis}
{Lewis} A.,  {Challinor} A.,  2006, \mn@doi [\physrep]
  {10.1016/j.physrep.2006.03.002}, \href
  {http://adsabs.harvard.edu/abs/2006PhR...429....1L} {429, 1}

\bibitem[\protect\citeauthoryear{{Meylan}, {Jetzer}, {North}, {Schneider},
  {Kochanek}  \& {Wambsganss}}{{Meylan} et~al.}{2006}]{kochanek}
{Meylan} G.,  {Jetzer} P.,  {North} P.,  {Schneider} P.,  {Kochanek} C.~S.,
  {Wambsganss} J.,  eds, 2006, {Gravitational Lensing: Strong, Weak and Micro}
  (\mn@eprint {} {astro-ph/0407232})

\bibitem[\protect\citeauthoryear{{Narayan} \& {Bartelmann}}{{Narayan} \&
  {Bartelmann}}{1996}]{narayan}
{Narayan} R.,  {Bartelmann} M.,  1996, ArXiv Astrophysics e-prints, \href
  {http://adsabs.harvard.edu/abs/1996astro.ph..6001N} {}

\bibitem[\protect\citeauthoryear{{Rusin} et~al.,}{{Rusin} et~al.}{2003}]{Rusin}
{Rusin} D.,  et~al., 2003, \mn@doi [\apj] {10.1086/346206}, \href
  {http://adsabs.harvard.edu/abs/2003ApJ...587..143R} {587, 143}

\bibitem[\protect\citeauthoryear{{Schneider}}{{Schneider}}{2003}]{schneider1}
{Schneider} P.,  2003, ArXiv Astrophysics e-prints, \href
  {http://adsabs.harvard.edu/abs/2003astro.ph..6465S} {}

\bibitem[\protect\citeauthoryear{{S{\'e}rsic}}{{S{\'e}rsic}}{1963}]{sersic1}
{S{\'e}rsic} J.~L.,  1963, Boletin de la Asociacion Argentina de Astronomia La
  Plata Argentina, \href {http://adsabs.harvard.edu/abs/1963BAAA....6...41S}
  {6, 41}

\bibitem[\protect\citeauthoryear{{Sersic}}{{Sersic}}{1968}]{sersic2}
{Sersic} J.~L.,  1968, {Atlas de galaxias australes}

\bibitem[\protect\citeauthoryear{{Somerville} \& {Dav{\'e}}}{{Somerville} \&
  {Dav{\'e}}}{2015}]{somerville}
{Somerville} R.~S.,  {Dav{\'e}} R.,  2015, \mn@doi [\araa]
  {10.1146/annurev-astro-082812-140951}, \href
  {http://adsabs.harvard.edu/abs/2015ARA%26A..53...51S} {53, 51}

\bibitem[\protect\citeauthoryear{{Vegetti} \& {Koopmans}}{{Vegetti} \&
  {Koopmans}}{2009}]{vegetti1}
{Vegetti} S.,  {Koopmans} L.~V.~E.,  2009, \mn@doi [\mnras]
  {10.1111/j.1365-2966.2009.15559.x}, \href
  {http://adsabs.harvard.edu/abs/2009MNRAS.400.1583V} {400, 1583}

\bibitem[\protect\citeauthoryear{{Vegetti}, {Koopmans}, {Bolton}, {Treu}  \&
  {Gavazzi}}{{Vegetti} et~al.}{2010}]{vegetti2}
{Vegetti} S.,  {Koopmans} L.~V.~E.,  {Bolton} A.,  {Treu} T.,   {Gavazzi} R.,
  2010, \mn@doi [\mnras] {10.1111/j.1365-2966.2010.16865.x}, \href
  {http://adsabs.harvard.edu/abs/2010MNRAS.408.1969V} {408, 1969}

\makeatother
\end{thebibliography}


\onecolumn

\appendix

\section{Detailed derivation of the two point correlation function}\label{APDPS}
Within the field of view of strong lensing we can apply flat sky approximation and thus we can expand the image intensity as follows
\begin{eqnarray}
I(\vx) &=& \int \dFT{\vk} I(\vk) e^{i\vk\cdot \vx} \nonumber \\
I(\vk) &=& \int \dFT{\vx} I(\vx) e^{-i\vk\cdot \vx}.
\label{Fourier}
\end{eqnarray}
If we assume the surface brightness fluctuations of image are statistically isotropic, the real space two point correlation function $\xi$ of the surface brightness can therefore only depend on the separation between the two points,
\begin{equation}
\langle I(\vx) I(\vx') \rangle = \xi^{II}(|\vx - \vx'|).
\end{equation}
With this assumption the covariance of the Fourier components of the surface brightness is
\begin{align}
\langle I(\vk) I^*(\vk') \rangle & = \int \dFT{\vx} \int \dFT{\vx'} e^{-i\vk\cdot\vx} e^{i\vk'\cdot\vx'} \xi^{II}(|\vx-\vx'|) \nonumber\\
& = \int \dFT{\vx} \int \dFT{\vr} e^{i(\vk'-\vk)\cdot\vx}e^{i\vk'\cdot \vr} \xi^{II}(r) \nonumber\\
& = \delta(\vk'-\vk)\int \ud^2\vr  e^{i\vk\cdot\vr} \xi^{II}(r).
\label{xi_relation}
\end{align}
In the second line we changed variables to $\vr = \vx -\vx'$ and then $\vr \rightarrow -\vr$, and have defined $r\equiv |\vr|$ which is the correlation length in image plane. The power spectrum of the surface brightness field of source is therefore diagonal in $\vk$, and is given by
\begin{equation}
\langle I(\vk) I^*(\vk') \rangle = P_{k}^{II} \delta(\vk-\vk').
\end{equation}
where we have defined the power spectrum $P_{k}^{II}$ as follows,
\begin{align}
P_{k}^{II} &= \int \ud^2\vr  e^{i\vk\cdot\vr} \xi^{II}(r)
\end{align}
Now if we use the following expansion of $e^{i\vk\cdot\vr}$ into Bessel functions $J_n(r)$ 
\begin{eqnarray}
e^{i kr\cos\phi} &=& \sum_{n=-\infty}^\infty  i^n J_n(kr) e^{i n \phi} \nonumber \\
&=& J_0(kr) + 2 \sum_{n=1}^\infty i^n J_n(kr)\cos(n\phi)
\label{BessJ}
\end{eqnarray}
and then if we integrate over $\phi$, the only term that remains is $J_0(r)$. This makes the Fourier transform as a Hankel transform which allows us to write the power spectrum as follows,
\begin{eqnarray}
P_{k}^{II} &=& \int \ud^2\vr  e^{i\vk\cdot\vr} \xi^{II}(r) \nonumber \\
&=& \int  r\ud r \int \ud\phi_\vr\, e^{i k r \cos(\phi_\vk -\phi_\vr)} \xi^{II}(r) \nonumber \\
&=& 2\pi \int r\ud r\, J_0(k r) \xi^{II}(r) 
\label{flatCl-corr}
\end{eqnarray}
If we neglect PSF, noise and window function the theoretical lensed correlation function $\xi^{II}(r)$ is given by,

\begin{eqnarray}
\xi^{II}(r) &=& \langle I(\vx) I(\vx') \rangle\nonumber \\
&=& \langle S(\vy) S(\vy') \rangle\nonumber \\
&=& \langle S(\vx - \nabla \psi_0 (\vx) -\nabla \delta \psi (\vx)) S(\vx' - \nabla \psi_0 (\vx') - \nabla \delta \psi (\vx')  ) \rangle \nonumber\\
&=& \int \dFT{\vk} \int \dFT{\vk'} e^{i\vk\cdot \vx} e^{-i\vk'\cdot\vx'} \langle e^{-i\vk\cdot \nabla \psi_0 (\vx)} e^{i\vk'\cdot \nabla \psi_0 (\vx')} \rangle  \nonumber \\ 
&& \qquad \qquad \qquad \langle e^{-i\vk\cdot \nabla \delta \psi (\vx)} e^{i\vk'\cdot \nabla \delta \psi (\vx')} \rangle \langle S(\vk)S(\vk')^*\rangle
 \nonumber\\
&=& \int \frac{\ud^2 \vk}{(2\pi)^2}\, P^{S} (k) e^{i \vk \cdot \vr} \langle e^{i\vk \cdot (\nabla \psi_0 (\vx') - \nabla \psi_0 (\vx))} \rangle
\langle e^{i\vk \cdot (\dalpha' - \dalpha)} \rangle_{\dalpha}.
\label{flat_lensed_corr_a}
\end{eqnarray}
where in the second and third lines above we have incorporated the principle of conservation of surface brightness and thereafter we have inserted the lens equation. In the fourth line we have expanded the source surface brightness in its Fourier modes without using any Taylor expansion or linear approximation where $\psi_0(\vx)$ and $\delta \psi(\vx)$ are the potentials for the smooth lens model and perturbations, respectively. Finally in the last line the power spectrum of the source is denoted by $P^{S}(k)$ and the deflection angle due to potential perturbations is denoted as $\dalpha = \nabla \delta \psi$.
Now if there are no perturbations, then Eq.\eqref{flat_lensed_corr_a} reduces to,

\begin{equation}
\xi^{II}(r)  = \int \frac{\ud^2 \vk}{(2\pi)^2}\, P^{S}(k) e^{i \vk \cdot \vr} \langle e^{i\vk \cdot (\nabla \psi_0 (\vx') - \nabla \psi_0 (\vx))} \rangle
\end{equation}
which basically turns into the two point correlation of the smooth model. Now using the standard result that the two point correlation function is Fourier transform of the power spectrum, we can write Eq.\eqref{flat_lensed_corr_a} as follows
\begin{equation}
\xi^{II}(r) = \int \frac{\ud^2 \vk}{(2\pi)^2}\, P_{s}^{II} (k) e^{i \vk \cdot \vr} \langle e^{i\vk \cdot (\dalpha' - \dalpha)} \rangle_{\dalpha}.
\end{equation}
where $P_{s}^{II} (k)$ is the power spectrum of the smooth lens model.

Now we have the following standard identity for a Gaussian variate $x$ with a complex source term:
\begin{equation}
\int^\infty_{-\infty} \ud x\, e^{-\frac{1}{2}ax^2 +iJx}=\sqrt{\frac{2\pi}{a}}e^{-J^2/2a}
\end{equation}
Using this, we get,
\begin{align}
\langle e^{ix} \rangle &= \frac{1}{\sqrt{2\pi}\sigma_x} \int^\infty_{-\infty} \ud x\, e^{ix}e^{-x^2/2\sigma_x^2} =e^{-\sigma_x^2/2}= e^{-\langle x^2\rangle/2}.
\label{gauss_avg}
\end{align}
So, if we assume $\dalpha$ is a Gaussian field, then $\vk \cdot (\dalpha' - \dalpha)$ is a Gaussian variate, and the expectation value in Eq.~\eqref{flat_lensed_corr_a} therefore reduces to
\begin{eqnarray}
\left\langle e^{i\vk \cdot (\dalpha' - \dalpha)} \right\rangle &=&
e^{- \frac{1}{2} \left\langle [\vk \cdot (\dalpha'-\dalpha)]^2 \right\rangle}
\label{expavg_a}
\end{eqnarray}
\section{The deflection angle structure function}\label{DASF}
\begin{eqnarray}
\langle \delta \alpha_i \delta \alpha_j' \rangle = A_1(r) \delta_{ij} + A_2(r) \hat{r}_i \hat{r}_j
\end{eqnarray}
To determine $A_1$ and $A_2$ we use two following properties of covariance matrix. First if we take trace of the covariance matrix then we get
\begin{eqnarray}
\langle \sdalpha_i \sdalpha_i' \rangle = 2 \, A_1(r) + A_2(r) \equiv \langle \dalpha \cdot \dalpha'\rangle  
\end{eqnarray}
and if the correlation matrix $\langle \alpha_i \alpha_j'\rangle$ is contracted with  $\hat{\mathbf{r}}_i\hat{\mathbf{r}}_j$ we get,
\begin{align}
\langle \sdalpha_i \sdalpha_j' \rangle \hat{\mathbf{r}}_i\hat{\mathbf{r}}_j = A_1(r) + A_2(r)
\end{align}
where we have used the Einstein's summation convention in both the equations above.

Now, from the theory of strong gravitational lensing, we know, 
\begin{equation}
\nabla \cdot \dalpha = 2\delta\kappa
\end{equation}
where $\kappa$ is the convergence or the dimensionless surface mass density corresponding to the lensing potential $\delta \psi$ of the dark matter substructure of the galaxy which was not incorporated into our previous smooth lens model and this is responsible for the deflections $\valpha$. Taking Fourier transform of the both sides of the above equation, we get
\begin{eqnarray}
ik_j \sdalpha_{j}(k)&=&2 \dkappa(k) \nonumber \\
\sdalpha_{j}(k)&=&\frac{2 \dkappa(k)}{ik_j}
\end{eqnarray}
Now, using the above relations and we get, 

\begin{eqnarray}
\langle \delta{\boldsymbol{\alpha}} \cdot \delta{\boldsymbol{\alpha}}' \rangle 
&=& \int \frac{\ud^2 \vk}{(2\pi)^2} \sdalpha_j^*(k')\sdalpha_j(k)  e^{i\vk\cdot \vr}  \nonumber\\
&=& 4 \int \frac{\ud^2 \vk}{(2\pi)^2} \frac{|\dkappa(k)|^2}{k^2} e^{i\vk\cdot \vr} \nonumber \\
&=& \frac{4}{2\pi} \int \frac{\ud k}{k} |\dkappa(k)|^2  J_0(kr) \nonumber \\
&=& 2 \, A_1(r) + A_2(r)
\end{eqnarray}

\begin{eqnarray}\label{deflec_contrct}
\langle \sdalpha_i \sdalpha_j'\rangle  \hat{\mathbf{r}}_i\hat{\mathbf{r}}_j &=& \hat{\mathbf{r}}_i\hat{\mathbf{r}}_j 
\int \frac{\ud^2 \vk}{(2\pi)^2} \sdalpha_i(k) \sdalpha_j^*(k')  e^{i\vk\cdot \vr}  \nonumber\\
&=& 4 \int \frac{\ud^2 \vk}{(2\pi)^2} (\hat{\mathbf{r}} \cdot \hat{\mathbf{k}})^2 \frac{|\dkappa(k)|^2}{k^2} e^{i\vk\cdot \vr} \nonumber\\
&=& 4 \int_0^\infty \frac{k \, \ud k}{(2\pi)^2} \frac{|\dkappa(k)|^2}{k^2} \int_0^{2\pi} \cos^2\phi \, e^{ikr\cos\phi} d\phi \, \nonumber \\
&=& 4 \int_0^\infty \frac{k \, \ud k}{(2\pi)^2} \frac{|\dkappa(k)|^2}{k^2} \int_0^{2\pi} d\phi\, \frac{[1 + \cos(2\phi)]}{2}\, e^{ikr\cos\phi} \, \nonumber \\
&=& \frac{1}{2} \, 4 \int_0^\infty \frac{k \, \ud k}{2\pi} \frac{|\dkappa(k)|^2}{k^2} (J_0(kr) - J_2(kr)) \nonumber \\
&=& A_1(r) + A_2(r) 
\end{eqnarray}
where we defined $\phi$ as the angle between $\vk$ and $\vr$, $\phi = \phi_\vk -\phi_\vr$ and used Eq.~\eqref{BessJ} to express the integrals in terms of Bessel functions. Now, comparing the last two results of the integrals we find,
\begin{eqnarray}
A_1(r) &=& \frac{1}{2} \Big( \frac{4}{2\pi} \int \frac{\ud k}{k} |\dkappa(k)|^2  J_0(kr) + \frac{4}{2\pi} \int \frac{\ud k}{k} |\dkappa(k)|^2  J_2(kr) \Big) \nonumber \\
&=& \frac{4}{2\pi} \int \frac{\ud k}{k} |\dkappa(k)|^2  \frac{J_1(kr)}{kr}
\end{eqnarray}
where we have used the following recursion relation of Bessel functions,
\begin{equation}
2 \, n \, \frac{J_n(x)}{x} = J_{n-1}(x) + J_{n+1}(x).
\end{equation}
And $A_2(r)$ turns out to be:
\begin{equation}
A_2(r)= - \frac{4}{2\pi} \int \frac{\ud k}{k} |\dkappa(k)|^2  J_2(kr)
\end{equation}

We now express the required expectation value in Eq.~\eqref{expavg_a} in terms of $A_1$ and $A_2$:
\begin{eqnarray}
&& \left\langle [\vk \cdot (\dalpha'-\dalpha)]^2 \right\rangle \nonumber \\
&=& k^i k^j\langle (\sdalpha_i'-\sdalpha_i)(\sdalpha_j'-\sdalpha_j)\rangle \nonumber\\
&=& k^i k^j \Big[ \langle \sdalpha_i \sdalpha_j \rangle + \langle \sdalpha'_i \sdalpha'_j \rangle - \langle \sdalpha'_i \sdalpha_j \rangle - \langle \sdalpha_i \sdalpha'_j \rangle \Big] \nonumber \\ 
&=& 2k^i k^j \Big[ \langle \sdalpha_i \sdalpha_j \rangle - \langle \sdalpha'_i \sdalpha_j \rangle \Big] \nonumber \\
&=& 2k^i k^j \Big[ \Big(  A_1(0) \delta_{ij} + A_2(0) \hat{r}_i \hat{r}_j \Big) - \Big( A_1(r) \delta_{ij} + A_2(r) \hat{r}_i \hat{r}_j \Big) \Big] \nonumber \\
&=& 2 \, k^2 [A_1(0) - A_1(r)] + 2k^i k^j \hat{r}_i \hat{r}_j [A_2(0) - A_2(r)]\nonumber\\
&=& 2 \, k^2 [A_1(0) - A_1(r)] + 2 k^2 \cos^2 \phi\, [ A_2(0) - A_2(r)] \nonumber \\
&=& k^2 \Big[ 2 \Big( A_1(0) - A_1(r) \Big) + \Big( A_2(0) - A_2(r) \Big) + \cos2\phi \Big( A_2(0) - A_2(r) \Big)\Big] \nonumber \\
&=& k^2 \sigma^2(r) + ( k_{r_{\parallel}}^2 - k_{r_{\perp}}^2 )\zeta(r)
\label{flat_expect}
\end{eqnarray}
\section{$A_1(r)$ and $A_2(r)$ in terms of hypergeometric function }\label{hgf}
Assuming a power law $\dkappa(k) \sim k^{-\gamma}$ we can write the values of $A_1(r)$ and $A_2(r)$ in terms of the generalized hypergeometric function ${}_{1}F_{2}$ and regularized generalized hypergeometric function ${}_{1}\tilde{F}_{2}$ using the following integral identities:
\begin{eqnarray}
\int k^{-\alpha} \frac{J_{n}(kr)}{kr} \ud k = \frac{2^{-n} \, k^{-\alpha}}{r(n-\alpha)\Gamma (n + 1)} \,  (kr)^n \, {}_{1}F_{2} \Big( \frac{n}{2} - \frac{\alpha}{2}; \frac{n}{2} - \frac{\alpha}{2} +1, n+1; -\frac{1}{4}k^{2}r^{2})
\end{eqnarray}

\begin{eqnarray}
\int k^{-\alpha} J_{n}(kr) \ud k = \frac{2^{-n} \, k^{1-\alpha}}{(-\alpha +n +1)\Gamma (n + 1)} \,  (kr)^n\, {}_{1}F_{2} \Big(- \frac{\alpha}{2} + \frac{n}{2} + \frac{1}{2};  - \frac{\alpha}{2} +\frac{n}{2} + \frac{3}{2}, n+1; -\frac{1}{4}k^{2}r^{2})
\end{eqnarray}

Here the generalized hypergeometric function ${}_{p}F_{q}$ is defined as follows:
\begin{eqnarray}
{}_{p}F_{q} (a_{1}, \dots a_{p}; b_{1}, \dots b_{q}; z)=\Sigma_{n=0}^{\infty}\frac{(a_{1})_{n} (a_{2})_{n} \cdots (a_{p})_{n}}{(b_{1})_{n} (b_{2})_{n} \cdots (b_{q})_{n}} \frac{z^{n}}{n!}
\end{eqnarray}
where we have used the following notation
\begin{eqnarray}
(x)_{n}=x(x-1) \cdots (x-n+1)=\frac{\Gamma(x+1)}{\Gamma(x-n+1)}
\end{eqnarray}

Using the above results we get,
\begin{eqnarray}
A_1(r) &=& \frac{4}{2\pi} \int \frac{\ud k}{k} |\delta \kappa(k)|^2  \frac{J_1(kr)}{kr} \nonumber \\
&=& \frac{4}{2\pi} \int k^{-\alpha} \frac{J_{1}(kr)}{kr} \ud k \nonumber \\
&=& \frac{4}{2\pi} \frac{k^{1-\alpha}}{2(1-\alpha)} \ {}_{1}{F}_{2} (\frac{1}{2}-\frac{\alpha}{2}; 2, \frac{3}{2}-\frac{\alpha}{2}; -\frac{1}{4}k^{2}r^{2})
\end{eqnarray}
where $\alpha = 2\gamma +1$. Similarly we can write $\Cgltwo(r)$ as follows:
\begin{eqnarray}
A_2(r) &=& - \frac{4}{2\pi} \int \frac{\ud k}{k} |\kappa(k)|^2  J_2(kr) \nonumber \\
&=& - \frac{4}{2\pi} \int k^{-\alpha} J_{2}(kr) \ud k \nonumber \\
&=& - \frac{4}{2\pi} \frac{1}{24- 8 \alpha} r^2 k^{3-\alpha} \ {}_{1}{F}_{2} (\frac{3}{2}-\frac{\alpha}{2}; 3, \frac{5}{2}-\frac{\alpha}{2};  -\frac{1}{4}r^{2}k^{2}) 
\end{eqnarray}

\section{Power spectrum of lens potential fluctuations and corresponding convergence maps}

\begin{figure*}
  \centering
{\includegraphics[width=145mm]{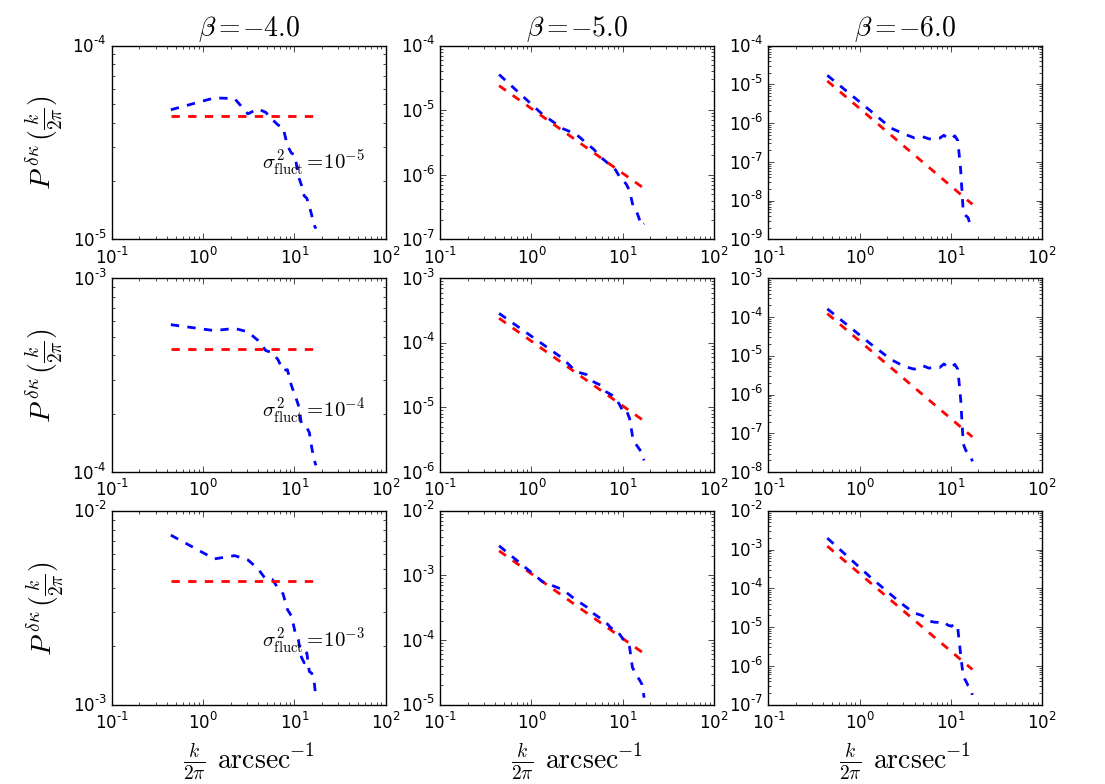}}
\caption{Theoretical (red) and numerically computed (blue) power spectra of $\delta \kappa$ (only one realization) corresponding to different combinations of variances and slopes of potential fluctuations, $\delta \psi$ as shown in Figure \ref{kappa}. The first, second and third row corresponds to variance levels of $10^{-5}, 10^{-4}$ and $10^{-3}$ and first, second and third column corresponds to slopes of -4.0, -5.0 and -6.0 respectively.}
\label{kappa_3}
\end{figure*}

\begin{figure*}
  \centering
{\includegraphics[width=145mm]{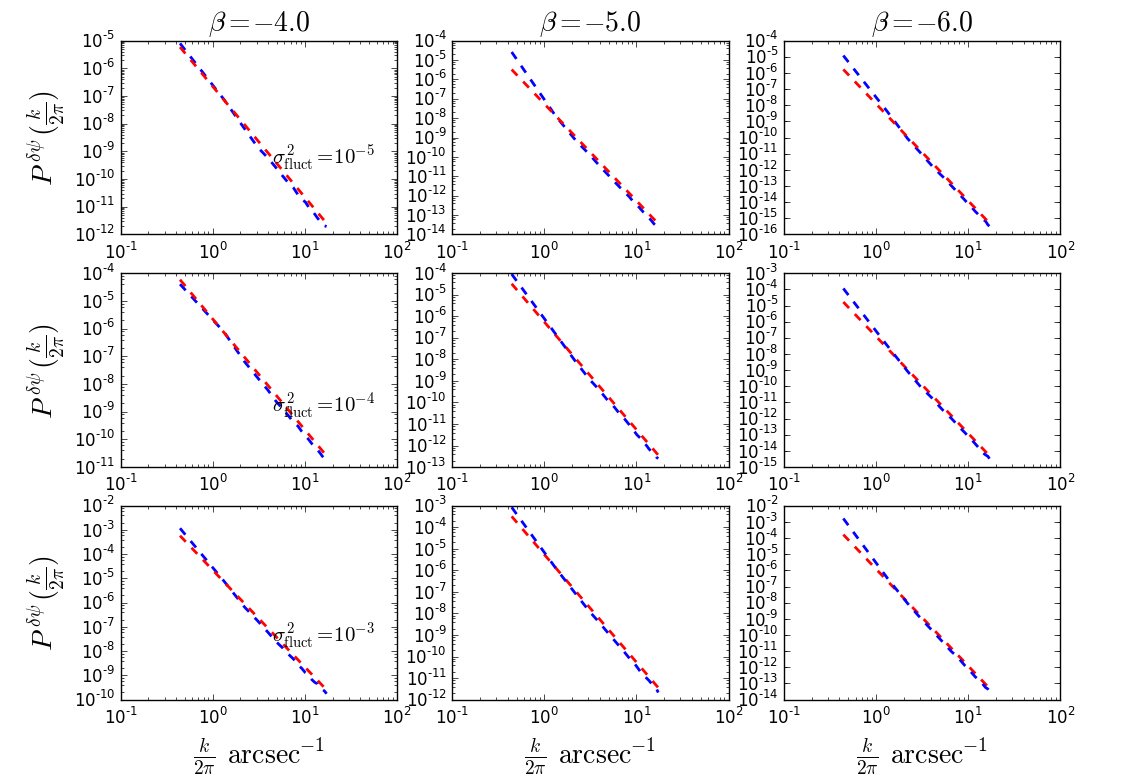}}
\caption{Theoretical (red) and numerically computed (blue) power spectra of lens potential fluctuations $\delta \psi$ corresponding to different combinations of variances $\sigma^{2}_{\rm{fluct}}$ and slopes $\beta$ shown in Figure \ref{kappa}, \ref{kappa_2} and \ref{kappa_3} . The first, second and third row corresponds to variance levels of $10^{-5}, 10^{-4}$, $10^{-3}$ and first, second and third column corresponds to slopes of -4.0, -5.0 and -6.0 respectively.}
\label{psi}
\end{figure*}

\begin{figure*}
  \centering
{\includegraphics[width=145mm]{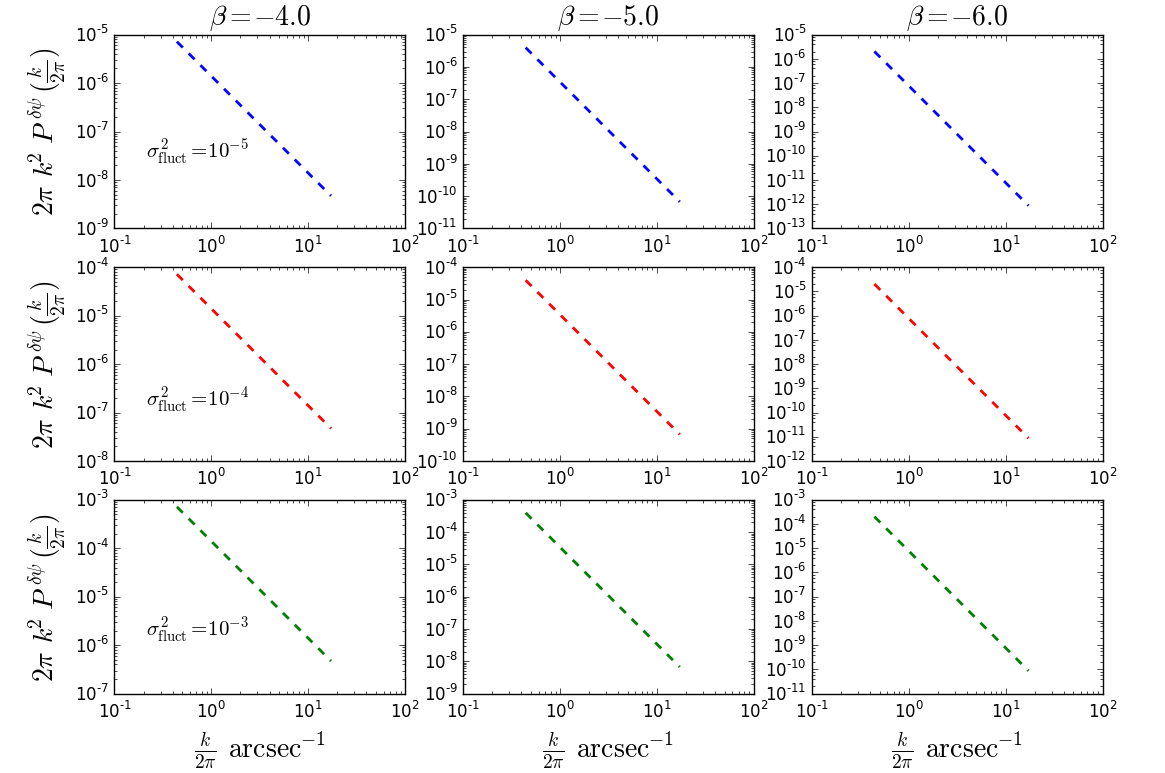}}
\caption{Plots of $2 \pi k^2 P^{\delta \psi}$ corresponding to different combinations of variances and slopes of potential fluctuations as shown in Figure \ref{psi}. }
\label{psi_2}
\end{figure*}


\bsp	
\label{lastpage}
\end{document}